# Advanced Spectroscopic Analyses on a:C-H Materials: Revisiting the EELS Characterization and its Coupling with multi-wavelength Raman Spectroscopy

L. Lajaunie[1‡], C. Pardanaud[2‡], C. Martin[2], P. Puech[3], C. Hu[4,5], M. J. Biggs[4,5], R. Arenal[1,6*]

[1]*Laboratorio de Microscopias Avanzadas (LMA), Instituto de Nanociencia de Aragon, Universidad de Zaragoza, 50018 Zaragoza, Spain*

[2]*Aix-Marseille Université,CNRS, PIIM, Marseille, France*

[3]*CEMES, Toulouse, France*

[4]*School of Chemical Engineering, The University of Adelaide, SA 5005, Australia*

[5]*School of Science, Loughborough University, Loughborough, LE11 3TU, UK*

[6]*Fundacion ARAID, 50018 Zaragoza, Spain*

\* *Corresponding author:* Tel: +34 976 762 985, E-mail address: arenal@unizar.es

[‡] These authors contributed equally

**Abstract**

Hydrogenated amorphous carbon thin films (a:C-H) are very promising materials for numerous applications. The growing of relevance of a:C-H is mainly due to the long-term stability of their outstanding properties. For improving their performances, a full understanding of their local chemistry is highly required. Fifteen years ago, electron energy-loss spectroscopy (EELS), developed in a transmission electron microscope (TEM), was the technique of choice to extract such kind of quantitative information on these materials. Other optical techniques, as Raman spectroscopy, are now clearly favored by the scientific community. However, they still lack of the spatial resolution offered by TEM-EELS. In addition, nowadays, the complexity of the physics phenomena behind EELS is better known. Here, a:C-H thin films have been isothermally annealed and the evolution of their physical and chemical parameters have been monitored at the local and macroscopic scales. In particular, chemical in-depth inhomogeneities and their origins are highlighted. Furthermore, a novel procedure to extract properly and reliably quantitative chemical information from EEL spectra is presented. Finally, the pertinence of empirical models used by the Raman community is discussed. These works demonstrate the pertinence of the combination of local and macroscopic analyses for a proper study of such complex materials.

**1. Introduction**

Hydrogenated amorphous carbon (a-C:H) thin films constitute a mix, at the local scale, of $sp^2$ clusters and $sp^3$ carbon domains with different contents of H. The physical and chemical properties of these films are strongly governed by on the balance between the $sp^2$ clustering and H defects [1,2]. The high stability of these materials makes them very suitable for a large range of applications including the use as protecting coatings for data storage devices in order to preserve both magnetic discs and recording heads from corrosion and wear [3–5]. Recently there is a renewed interest for these materials as they exhibits a superlow friction coefficient and long wear life in ultra-high vacuum which makes them one of the most promising solid lubricant coating candidate for aerospace applications [6–10]. The origin of the superlow friction is still under investigation but seems to be closely related to the passivation of the surface by the hydrogen and to the $sp^3$ to $sp^2$ rehybridization [8–11]. The proper characterization of the hydrogen content together with the $sp^2$ fraction is thus of particular importance. These coatings need also to be thermally stable and changes in a-C:H local chemistry during thermal annealing have then regained an interest these last few years [3,4,12–22].

Due to the extremely good sensitivity on structural modifications involving $sp^2$ carbons, Raman spectroscopy analysis has been associated to most of the last studies on a:C-H materials [3,4,17,19,23–30]. The reason behind this high sensitivity is the resonance phenomena between the laser wavelength excitation used and a local band gap of a given $sp^2$ cluster in the amorphous carbon. As $sp^2$ clusters in amorphous carbon display inhomogeneous structural and topological disorders at the nanometric scale, their band gaps are modified subsequently [31]. Then, Raman spectra of amorphous carbons depend on clustering of the $sp^2$-phase, bond length and bond-angle disorder, presence of $sp^2$ rings and chains, $sp^2/sp^3$ ratio and inhomogeneity [23,24,27]. The optical band gap for amorphous carbon films is inversely related to the aromatic cluster size and the disorder potential of the samples [2,31,32]. Because of that, multi-wavelength Raman spectroscopy has been used on various $sp^2$ carbon containing materials to obtain information on the structure ranging from amorphous carbons or soots, to graphene or carbon nanotubes [23–26,28,33–40]. However, Raman spectroscopy presents some limitations for getting directly this chemical information from these materials at the local scale. Thus, in addition of this spectroscopic technique, electron energy-loss spectroscopy (EELS) performed in a (Scanning) transmission electron microscope ((S)TEM) is a complementary technique, which gives access to a wealth of structural, chemical and physical information with nanometer to sub-angstrom spatial resolution [41–47]. Analysis of core-loss EELS spectra yields qualitative and quantitative information on the elemental composition and on the local chemical environment (bonding configuration, valence state…) [48–52]. The fine structures of the C-K edge can be used to distinguish the different carbon allotropes [41,45,48,53–55]. The sharp peak at 285 eV corresponding to transitions to $\pi^*$ states (Fig. 1) is observed in the graphite spectrum but not in the diamond spectrum, the transitions to $\sigma^*$ states appearing at higher energy (around 292 eV) [48,56]. Beyond this qualitative description, the $sp^2$ fraction can also be determined from thorough analysis of the C-K edge [57–60]. However, the determination of the $sp^2$ fraction of carbon-based materials from core-loss EELS is not an easy task and two main pitfalls should be avoided. The first one lies in the modeling of the $\pi^*$ and $\sigma^*$ components of the C-K edge spectrum (*i.e.* the determination of the *R ratio*) and the second one concerns the structural and electronic anisotropy of these $sp^2$ carbon materials. Regarding this first limitation, most of the models in the literature use an improper combination of Gaussian and/or Lorentzian functions [61,62], whereas *ab-initio* calculations have shown that the $\pi^*$ component is asymmetric and spreads out several dozens of eV after the edge threshold [60,63]. As a direct consequence, models using a combination of Gaussian and/or Lorentzian functions need to add another artificial component to compensate the real asymmetry of the $\pi^*$ character [61,62,64]. Others methods based on the integration of two-windows are inappropriate from a chemical point-of view as they do not consider all the transitions present in the EELS spectra and thus failed to take into account the contribution of heretrospecies [65,66]. The next step to determine the $sp^2$ fraction requires the determination of the $\pi^*$ and $\sigma^*$ spectral weights from a reference sample showing 100% $sp^2$ bonding and in the same experimental conditions [58,67]. Graphite is then the ideal reference sample. However, it is an anisotropic material and the second pitfall lies in the sensitivity of the EELS spectra to the anisotropy of the material [45,46,55,68–73].

The spectral weight of the $\pi^*$ and $\sigma^*$ components of anisotropic carbon-base materials depends then strongly on the specimen orientation and on the convergence and collection angles [68–70,74]. A particular setup of convergence and collection angles can be used to alleviate the dependence of the C-K edge spectrum with the specimen orientation (the so-called *magic-angle condition*) [68–72,75]. The collection angle at magic angle condition for an energy-loss of 290 eV should be 4.9 mrad at 80 kV, 1.2 mrad at 200 kV and 0.7 mrad at 300 kV. At the highest voltage, the magic angle condition for carbon is then nearly impossible to attain while at lower voltage, in particular in STEM mode, a larger collection angle is needed to ensure a good signal/noise ratio. Fifteen years ago, according to the seminal work of Ferrari et al. [57], EELS was technique of choice to get quantitative information on the local chemistry of a-C:H materials. Nowadays, our knowledge of the physics phenomena behind the EEL spectroscopy has considerably improved, but surprisingly, this technique is now regarded as time-consuming and difficult to properly interpret. Under these circumstances, the community has favored the use of Raman spectroscopy. In the present work, we will show the pertinence of EELS and the coupling of this technique with Raman spectroscopy for obtaining the maximum information of these materials. In particular, the goals of these studies are, not only to get more insight on the local chemistry of a-C:H thin films, but also to provide some guidance regarding their characterizations by EELS and Raman spectroscopies.

In this paper, we investigated the modifications followed by a-C:H thin films on silicon substrate during long-term isothermal annealing periods, up to 2500 min. This paper is divided as follows. In section 2, the Raman and EELS methodologies are presented and a novel method to determine the $sp^2$ fraction from core-loss EELS is deeply described. A general overview of the Raman and EELS spectra are given in section 3. The evolution of physical and chemical parameters as function of the annealing time are discussed from a macroscopic point of view in section 4, together with their evolutions as a function of the distance between the electron probe and the substrate. Finally, the pertinence of empirical models used by the Raman community for quantitatively linking the sp$^2$ fraction with the evolution of Raman parameters [29,30] will be reviewed. To sum up, these works demonstrate and illustrate the perfect coupling of these local and macroscopic analyses for a proper study of such complex materials.

## 2. Materials and methods

### 2.1 Materials

A hard amorphous, hydrogenated carbon film (a-C:H), with a thickness around 300 nm and an initial hydrogen content (H/H+C) of 30 at. %, was deposited on a Si wafer on the driven electrode of a capacitively coupled radio frequency plasma (13.56 MHz) in pure methane at 2 Pa and a DC self-bias of -200V [17,76]. To vary the hydrogen content and structure, the layer was cut in several samples that were heated under 1.5 bar argon atmosphere at 500°C during t = 2, 15, 60, 120, 500, 1000 and 1500 minutes. For Raman spectroscopy analysis, a set of reference samples (heated from room temperature to 1000 °C under ultra-high vacuum, with H content measured by ion beam analysis and described in detail in [19]) has also been used for comparison. The attenuated total reflectance (ATR) infrared spectra of the different samples were taken with a Bruker IFS 66/S spectrometer, which was purged with dry air. The samples were used without further purification and they have been ground before measurement. The ATR spectra were recorded using a diamond or a germanium ATR crystal and a deuterated triglycine sulfate detector. A resolution of 2 cm was used.

TEM samples were prepared in the cross-section geometry; they were mechanically thinned using a tripod polisher down to 10 µm, and then ion-milled in a GATAN-PIPS apparatus at low energy (2.5 keV, Ar) and low incidence (± 8°) to minimize irradiation damage. It should be noted that ionic irradiation of amorphous carbon at such low energy should not alter significantly the surface of the sample [77]. Highly ordered pyrolytic graphite (HOPG) samples were also prepared for EELS analysis by ultrasonic vibration in pure ethanol during ten minutes.

**2.2 Multi-wavelength Raman set-ups and data treatment**

Multi-wavelength Raman spectra, with laser wavelength $\lambda_L$ = 633, 514, 407, 325 and 266 nm, were obtained at room temperature, using three different set up. For $\lambda_L$ = 633, 514 and 325 nm, Raman spectra were obtained in a Horiba-Jobin-Yvon HR LabRAM spectrometer. For $\lambda_L$ = 407 nm, Raman spectra were obtained in a T64000 Horiba-Jobin-Yvon apparatus. For $\lambda_L$ = 266 nm, Raman spectra were also obtained with a Horiba-Jobin-Yvon HR Labram spectrometer. The objective lens was a 40x magnification. The detector was a Horiba Jobin-Ivon Synapse charged coupled device detection system. To avoid irradiation damage at 266 nm excitation wavelength, a ~ 1 $\mu m^2$ laser spot was scanned over the sample on a 60x60 $\mu m^2$ area has been employed. The scanning time was 1 second per spectrum. We applied a similar procedure for the $\lambda_L$ = 325 nm radiation. The laser power was chosen sufficiently low to prevent damaging of the samples (roughly 1 mW$\mu m^{-2}$ on the sample) but sufficiently high to have a good signal to noise ratio.

The main Raman parameters analyzed were: the G peak wavenumber ($\sigma_G$) and its full-width at half-maximum ($\Gamma_G$) the relative heights of the G and D bands ($H_D/H_G$) and the m/$H_G$ ratio, where m is the slope of the photoluminescence background of the spectra. Concerning the different parameters, $\sigma_G$ gives information on the structural disorder and/or clustering, $\Gamma_G$ gives information on the topological disorder, $H_D/H_G$ and m/$H_G$ gives mainly information on the H-content [3,19]. A linear background was subtracted and heights $H_D$ and HG were measured without any fitting to prevent from ambiguousness due to a model-dependent fitting procedure [18,19]. This last choice was done because a debate still exists in the literature concerning how to fit properly amorphous carbon Raman spectra [1,30,78,79]. $H_D$ was therefore measured at its apparent maximum, except when the D band maximum was not well enough defined. For these samples $H_D$ was taken arbitrarily at 1370 cm$^{-1}$.

**2.3 EELS acquisition and data treatment**

EELS spectra were acquired on a FEI Tecnai F30 operating at 300 kV and equipped with a Gatan Tridiem 863 spectrometer. Convergence and collection angles were 6.1 and 5.9 mrad, respectively. To minimize carbon contamination and irradiation beam damage, experiments were performed at liquid nitrogen temperature. For the same reasons, the probe, of 1-2 nm, was constantly rastering a square area of 20 nm × 20 nm during the EELS acquisition. The rastering speed was 2 μs/pixel. The size of the rastered areas was much larger than sp$^2$ clusters size in amorphous carbons (typically a few nanometers [27]), which guarantees an EELS response independent of the orientation of the sp$^2$ clusters. The energy resolution, measured as the full width at half maximum of the zero loss peak (ZLP), was 1.2 eV with an energy dispersion of 0.2 eV/pixel and 0.9 eV for a dispersion of 0.05 eV/pixel. The first dispersion was used for carbon and oxygen elemental quantification and the second one for C-K edge fine-structures analysis and determination of the sp$^2$ fraction. The typical acquisition times for C-K edge and low-loss spectra were 20 and 0.2 seconds, respectively. At least 15 spectra per sample were acquired and processed to check the homogeneity of the samples. The spectra were also acquired at different positions from the Si substrate. Background subtraction for the C-K edge was performed by modeling the usual inverse power law function and the multiple scattering was then removed by Fourier-ratio deconvolution [41] with the low-loss spectrum obtained for exactly the same region of the sample. The maximum of the $\pi^*$ peak was fixed at 285 eV to minimize systematic error from peak position.

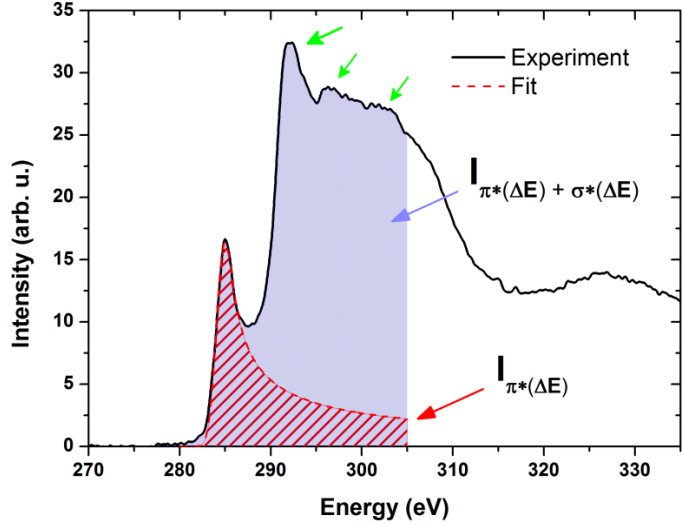

**Fig. 1** EELS spectra of a graphite sample illustrating the method used to determine the *R* ratio (see text for more details). The green arrows highlight the fine structures of the graphite spectra. (A color version of this figure can be viewed online)

The first step in the determination of the sp$^2$ fraction requires the calculation of the *R* ratio which is defined $R = I_{\pi^*(\Delta E)}/I(_{\pi^*(\Delta E)+\sigma^*(\Delta E)})$, where $I_{\pi^*(\Delta E)}$ and $I_{\sigma^*(\Delta E)}$ are the spectral weights of the transitions to the $\pi^*$ and $\sigma^*$ states, respectively. The method used in this work for the determination of *R* is illustrated in Fig. 1 and took inspiration from the work of Bernier *et al.* [60] in which the $\pi^*$ peak is modeled by using a sum of Gaussian functions of equal standard deviation and separated by 0.5 eV. Only three unknown parameters are required for this fit, the amplitude of all the Gaussian functions being constrained to match the decrease of $\pi^*$ contribution in the density of states of graphite. This method pursues thus the same goal than the one proposed by Titantah *et al.* [63]: to reproduce the $\pi^*$ character of the graphite C-K edge which, as highlighted by *ab-initio* calculations [60,63], is asymmetric and go on several dozens of eV after the edge threshold and thus overlap with the $\sigma^*$ states. This method has also been applied successfully to amorphous carbon with various sp$^2$ ratios [60]. The fit was performed between 284 and 286 eV in order to reproduce precisely the shape of the $\pi^*$ peak and the modeled function was then extended toward lower and higher energies. The integrated intensity of the $\pi^*$ character $I_{\pi^*(\Delta E)}$, was then determined by area integration of the modeled curve from 282 to 305 eV (red hatched area in Fig. 1). The integrated intensity $I_{(\pi^*(\Delta E)+\sigma^*(\Delta E))}$, corresponding to the summation of the contributions of the $\pi^*$ and $\sigma^*$ characters, was then determined by area integration of the experimental spectrum from 282 to 305 eV (blue area in Fig. 1) in order to calculate the $R$ ratio. The precision of this method was checked on HOPG platelets priory orientated by electron diffraction and tilted away from the basal plane by the same amount of degree (54.7°). The choice of this angle will be explained in the following. This yields an $R$ ratio of $0.21 \pm 0.02$ (less than 1 % of standard errors) and constitutes thus an improvement of the method of Bernier *et al.* [60] in which the $\sigma^*$ contribution is modelled by a Gaussian function centered at 292 eV and which yields a standard error of 4 % [60]. In addition, this method can be easily modified to take into account contributions from heterospecies (see Fig. S1 in the Supplementary Data). This has been tested on oxidized-graphenic nanoplatelets sample which contain 12 % of oxygen [80]. In this case, the error on the determination of the $R$ ratio was estimated to be equal to 1.6 %. It is slightly larger than the original method due to the presence of one additional parameter. This modified method has been used for the samples annealed 1000 and 2500 minutes (section 4.2).

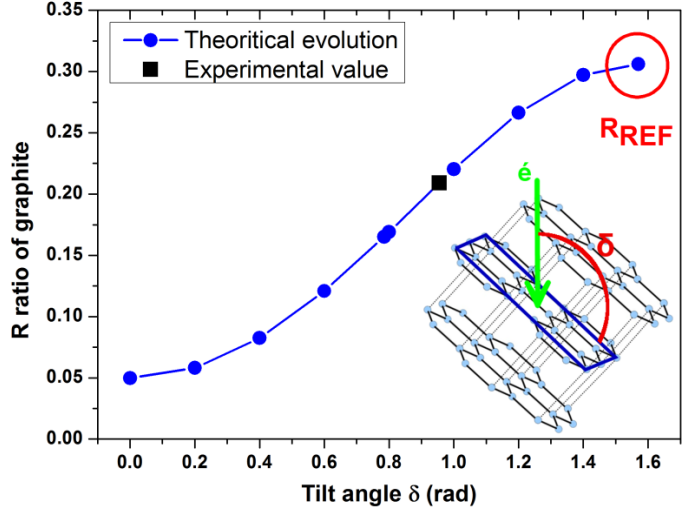

**Fig. 2** Variation of the *R* ratio as a function of the tilt angle, $\delta$. $\delta$ is the angle between the graphite basal plane and the incident electron beam. (A color version of this figure can be viewed online).

Once the *R* ratio was determined, the $sp^2$ fraction of the amorphous carbons was then calculated by using the following relation $sp^2 \% = R/R_{REF}$ [67], where $R_{REF}$ is the *R* ratio obtained from a reference showing 100% $sp^2$ bonding. This work does not use magic angle condition because, as abovementioned, such condition at 300 kV is nearly impossible to attain, and the obtained results show that this condition is not required, see below. $R_{REF}$ was then chosen as the maximum *R* value that could be obtained from an HOPG sample in the same experimental condition. For this purpose, a specific routine in Matlab was written based on the analytical expression of Bocquet *et al.* [75], which describes the variation of *R* as a function of all the experimental settings (convergence angle, collection angle, tilt angle and acceleration voltage) for uniaxial materials. This expression is based on relativistic calculation of the differential scattering cross-section and its main advantage is that it needs only one experimental value of *R* to derive *R* for any other experimental settings. Thanks to this routine, the variation of *R* as a function of the tilt angle (labeled as $\delta$ in the following) for graphite was derived by using the $R$ ratio experimentally determined from HOPG platelets tilted by 54.7° (Fig. 2). According to the literature, the $R$ value determined at this angle coincides with the one that would be obtained in magic angle condition [68,75]. In addition, this angle, which is nearly the maximum angle experimentally that can be obtained, has been chosen to minimize the uncertainty on the determination of $R$, as it lies close to the center $R=f(\delta)$ curve. As it can be seen from Fig. 2, the maximum value of $R$ is obtained for $\delta = \pi/2$ *rad i.e.* when the basal planes of graphite are parallel to the electron beam (a configuration which is experimentally impossible to obtain) and this value was used as $R_{REF}$ to derive the $sp^2$ fraction. It should be noted that this method to determine the $sp^2$ fraction is valid only if the $sp^2$ clusters of the a:C-H material do not show any preferential orientation at the scale of the apparent electron probe during the EELS acquisition (20 nm × 20 nm in this work). This has been tested by checking that there is no variation of $sp^2$ fraction as a function of tilt angle for the amorphous carbon samples. Finally the most representative core-loss spectra were submitted to the open-access EELS Database as references [81].

In addition to core-loss spectra, low-loss spectra were also acquired to determine the mass density of such samples. For this purpose, the low-loss spectra were deconvoluted by the ZLP by using the PEELS program [82] and the single-scattering spectra were obtained following Stephen's procedure [83]. For each spectrum, the ($\pi$ + $\sigma$) volume plasmon was then modeled over the top 75% of the plasmon peak intensity in order to extract the plasmon energy, $E_P$ (see Fig. S2 in the Supplementary Data). By assuming that carbon contributes to four valence electrons, oxygen to six and hydrogen to one, the following relationship was then derived by linking the mass density with the valence electron density [41,57]:

$$\rho = \frac{E_P^2 \times m^* \times \epsilon_0 \times \mu \times (12X_C + 16X_O + X_H)}{\hbar^2 \times e^2 \times (4X_C + 6X_O + X_H)} \quad \text{(Eq. 1)}$$

Where $\rho$ is the mass density, $E_P$ is the plasmon energy determined by the Drude model, $\varepsilon_0$ the vacuum permittivity, $\mu$ the atomic mass unit, $e$ the elementary charge, $\hbar$ the reduced Planck constant, $X_H$ the hydrogen fraction determined by Raman spectroscopy, $X_O$ the oxygen fraction determined by EELS (and by assuming that $X_C + X_O + X_H = 1$) and $m^*$ the effective electron band mass taken as $m^* = 0.87\,m$ [57,59,84].

## 3. RESULTS

### 3.1 Raman and EELS spectra

Raman spectroscopy has been used for getting different information, as for retrieving the H-content, as the $H_D/H_G$ ratio of a-C:H layers can be related to the H-concentration better than the m/$H_G$ ratio [19]. In the Fig. S3 of the Supplementary Data, the calibration method used to determine the hydrogen content by monitoring the evolution of the $H_D/H_G$ for a set of reference samples using wavelengths of 407, 514 and 633 nm is detailed. Below the variation of the other spectroscopic parameters of the samples will be discussed as they give complementary qualitative information and trends about disorder in the material.

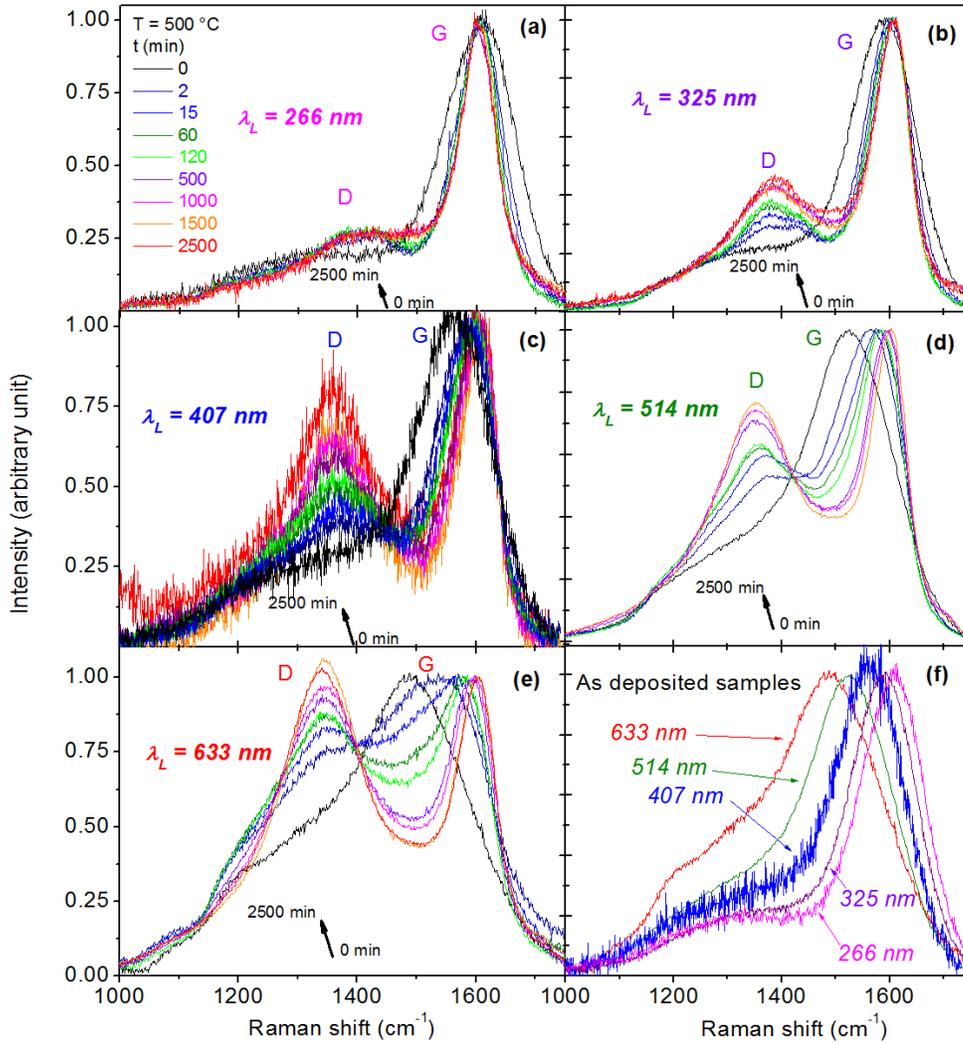

**Fig. 3** Raman spectra of the a-C:H samples heated at 500°C during 0 to 1500 minutes. (a) Spectra using a $\lambda_L$=266 nm laser. (b) Spectra using a $\lambda_L$=325 nm laser. (c) Spectra using a $\lambda_L$=407 nm laser. (d) Spectra using a $\lambda_L$=514 nm laser. (e) Spectra using a $\lambda_L$=633 nm laser. (f) As-deposited samples using all the lasers. (A color version of this figure can be viewed online)

Fig.3 displays the Raman spectra of the a-C:H samples heated at 500°C during 0 to 2500 minutes under argon atmosphere. From Fig. 3a to Fig. 3e these spectra are displayed increasing wavelength of the laser used from 266 nm to 633 nm, respectively. Due to the resonance mechanism occurring between the incident light and the band gap produced by sp$^2$ atoms embedded in small aromatic clusters, spectra of a given sample looks different depending on the wavelength of the laser used [85]. Two general trends on all these sub-figures are: first, all the spectra contain two main bands (the G and D bands of amorphous carbons), second, when time of heating increases, the D over G band ratio increases. This latest effect is associated to a bandwidth decrease. Except for Fig. 3a, where a redshift of $\sigma_G$ is observed with time, the increase and decrease with time of respectively $H_D/H_G$ and $\Gamma_G$ are also accompanied by a $\sigma_G$ blue-shift, as reported first in [86] for heated amorphous carbons. Fig. 3f displays the spectra of the as deposited samples, showing a blue-shift and narrowing of the G band and a decrease of the $H_D/H_G$ ratio with the decrease of the laser wavelength, as expected from [26]. The shape of all these spectra gives qualitative information: the isothermal heating produces structural changes, but the level of disorder remains high in the material after heating, as the D band is still present. All the physical implications about the variation of the spectroscopic parameters are discussed below.

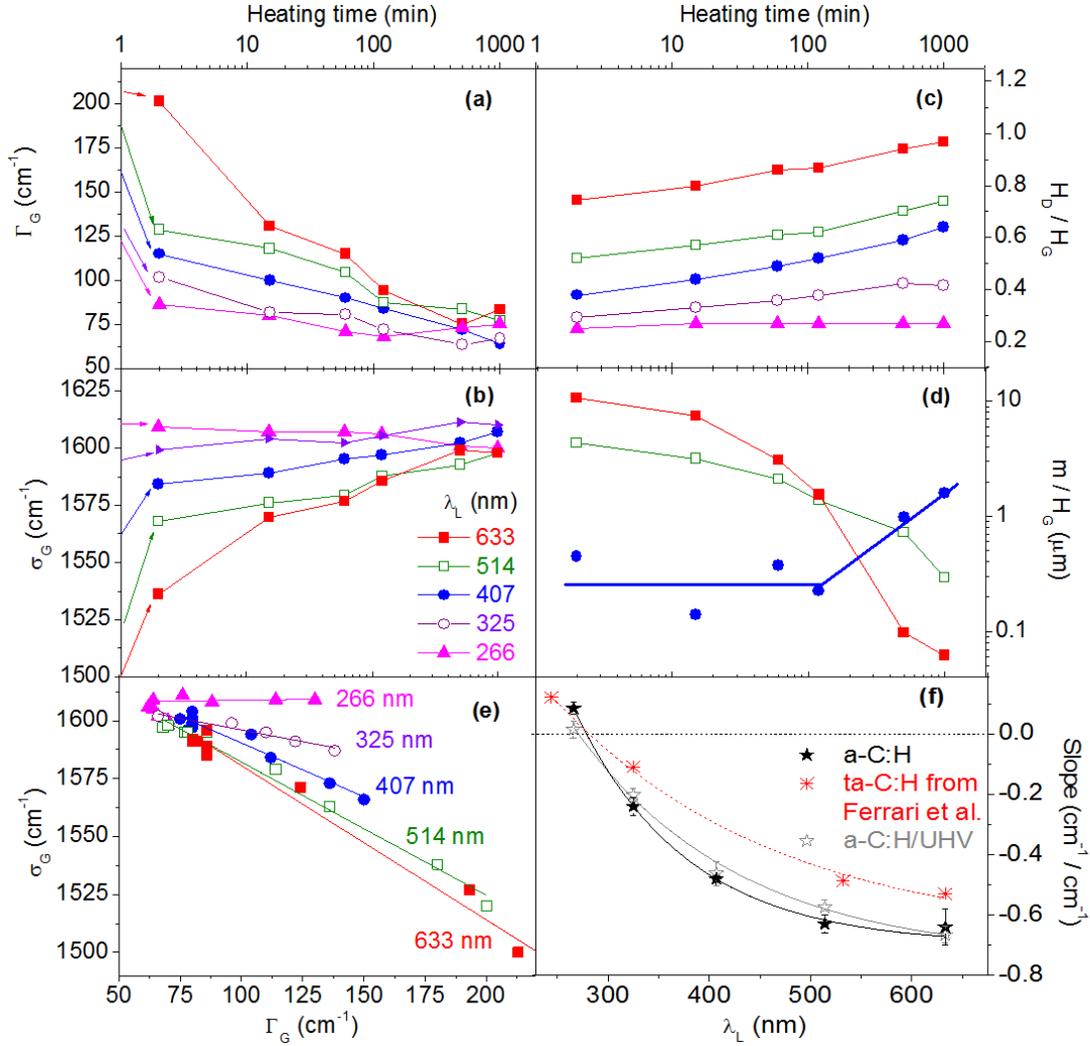

**Fig. 4** –Time evolution of apparent Raman parameters of the a-C:H heated at 500°C. Labels of sub figures *a*, *b*, *c* and *d* are the same and displayed only in *4b*. The origin of each arrow in *4a* and *4b* displays the "as deposited value" at heating time = 0 min. (a) G band width $\Gamma_G$, (b) G band position $\sigma_G$, (c) D and G band height ratio $H_D/H_G$ as a function of the annealing time. (d) $m/H_G$ parameter, m being the slope of the photoluminescence background. Multi-wavelength $\sigma_G(\Gamma_G)$ plot. (e) Plot for samples heated at 500°C under argon atmosphere during 2 to 1500 minutes. (f) Slopes extracted from the curves of *4e*. The ta-C:H measurements are taken from [26]. (A color version of this figure can be viewed online)

Fig. 4 displays the evolution of the apparent Raman spectroscopic parameters of the a-C:H sample heated at 500°C during 2 minutes to 2500 minutes. $\Gamma_G$, $\sigma_G$, and $H_D/H_G$ measured for the five laser wavelengths are displayed in Fig. 4a, 4b and 4c, respectively. The values of $m/H_G$ measured for the three lowest laser wavelength are plotted in Fig. 4d. For the other wavelengths, slopes are close to zero. The time evolution of these parameters is used in Fig. 4e and 4f to derive the H content [19,20,36] and the dispersion curve of the G band [26,27]. In Fig. 4a and 4b $\Gamma_G$ and $\sigma_G$, whatever the laser wavelength used, evolve rapidly between the as deposited sample and 2 minutes of heating, and then evolve monotically from 2 minutes to 1500 minutes, converging to a unique value close to $\Gamma_G \approx 60\text{-}70$ cm$^{-1}$ and $\sigma_G \approx 1600\text{-}1605$ cm$^{-1}$. This implies that the aromatic domain size at the end of the isoterm is a few nanometer large [27]. In addition, the convergence of the five $\sigma_G$ curves means the small aromatic clusters leading to the resonance effect have disappeared from the sample in 1000 minutes at 500°C [85]. The slopes in Fig. 4b are more pronounced for the higher wavelengths used.

In Fig. 4c, $H_D/H_G$ increases monotically with heating time for $\lambda_L$=633, 514 and 407 nm. $H_D/H_G$ is constant for 266 nm and saturates after 500 minutes for 325 nm. The difference of behavior for UV and visible light is due to a strong enhancement of the UV Raman cross-section of the G band with the H content [39]. The effect of this enhancement can also be seen on the reference sample that has been heated under UHV conditions (see Fig. S3 in SI) and it implies that for wavelengths shorter than 407 nm (*i. e.* 325 and 266 nm here), $H_D/H_G$ cannot be used to estimate the H-content. For 633, 514 and 407 nm, we use the linear relation existing between the H-content and $H_D/H_G$ to determine the H-content evolution with time (see Fig. S3 in Supplementary Data).

Fig. 4(d) shows that m/$H_G$ decreases for $\lambda_L$=633 and 514 nm and the opposite trend is observed for $\lambda_L$=407 nm. For $\lambda_L$=633 and 514 nm, the decrease has been interpreted as being due to the release of hydrogen previously bonded to a $sp^3$ carbon atom [19]. Furthermore, the evolution of m/$H_G$ does not evolve linearly with the log of time whereas it is the case for $H_D/H_G$. This difference of behavior suggests that H bonded to $sp^3$ carbon is released more easily than H bonded to $sp^2$ carbon [19]. The increase of m/$H_G$ for t > 100 minutes using $\lambda_L$=407 nm shows that there is an increase of photoluminescence in the material. The interpretation of this result is not clear yet, but it could be due to an experimental bias as a result of the change of the absorption coefficient of the layer with the wavelength: whereas the Raman signature of the underlying silicon wafer is seen during all the kinetic experiments carried out at 633 and 514 nm (not shown here but proving the laser probes all the thickness), it is not the case for shorter wavelengths, meaning Raman spectroscopy only probes a part of the sample far from the silicon/carbon interface at these wavelength. As it will be shown later, the surface starts to be polluted by oxygen for 1000 and 2500 minutes so that the m/$H_G$ increase can be due to photoluminescence created by oxygen impurity close to the surface. This last point is supported by the fact that such a m/$H_G$ with $\lambda_L$=407 nm has not been observed for the reference sample that has been heat treated under UHV conditions (see Fig. S3 in the Supplementary Data).

Fig. 4e displays the comparison of multi-wavelength $\sigma_G(\Gamma_G)$ plots samples heated at 500 °C during 2 to 1500 minutes under argon atmosphere. The behavior is the same for all the different excitation wavelengths (except for 266 nm): increasing time results in moving from the down right corner to the up left corner. The same trends are observed when plotting $\sigma_G$ in function of $\Gamma_G$ for the reference samples that have been heated under UHV conditions up to 1000°C (not shown here). Thus, it means that the two spectroscopic parameters, $\sigma_G$ and $\Gamma_G$, independently of the annealing method, are related each other. Slopes corresponding to all the wavelength are compared in Fig. 4f with that of ta-C:H sample (data taken from [26]), known to contain more $sp^3$ carbons than a-C:H sample. The evolution of this slope with $\lambda_L$ is more pronounced for the a-C:H sample than that for the ta-C:H sample. The origin of that may be due to the difference in the coupling between $sp^2$ and $sp^3$ carbons in a-C:H and ta-C:H, but it as to be tested on other materials.

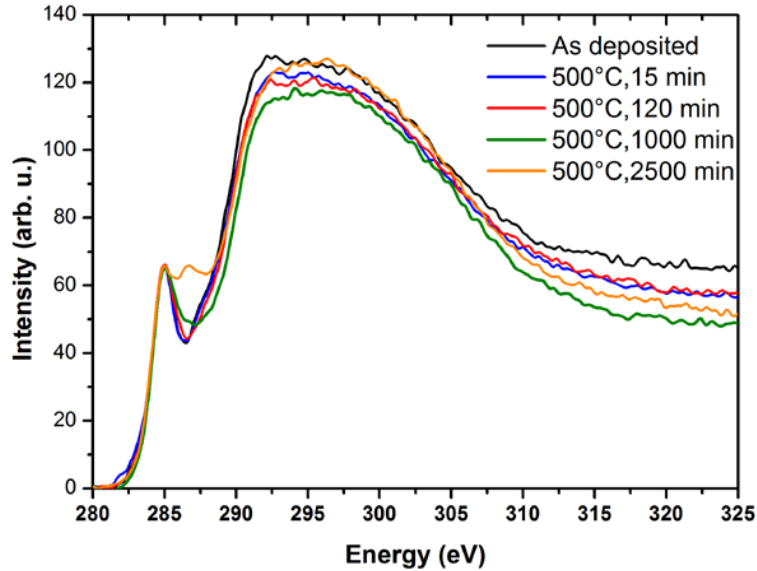

**Fig. 5** –C-K edge energy-loss near edge fine structures for the as-deposited and annealed samples. (A color version of this figure can be viewed online)

Fig. 5 shows the most representative EELS spectra for the as-deposited and annealed samples. All these spectra show similar fine-structures: the π* peak is clearly visible around 285 eV and the massif between 290 and 305 eV is nearly featureless. This is a typical signature of amorphous carbons [23,48]. The main striking feature of Fig. 5 lies in the intensity of the massif between 290 and 305 eV, which differs from sample to sample and which clearly highlights a slight variation of the $sp^2$ fraction. Indeed, to give a qualitative indicator of the $sp^2$ fraction, the spectra have been normalized for displaying the same maximum intensity of the π* peak. In addition, samples with the longest annealing time do not show the typical fine-structures of a perfect, pure and high quality $sp^2$ material (green arrows in Fig. 1) and then, it can be excluded that a full graphitization process occurred during the annealing. However a more detailed analysis (see below) shows the real different variations between all these samples, which is much more complex than the one displayed in Fig. 5. In particular, variations of the fine-structures near of the edge (ELNES) are observed for the as-deposited sample and samples annealed during 1 000 and 2500 minutes. These changes are related to chemical inhomogeneities, having different origins. The most significant change between them is the supplementary peak around 287 eV for the sample annealed during 2500 min which leads to an increase of the intensity between π* peak and the σ* massif starting around 292 eV. All these chemical inhomogeneities will be discussed in detail in Section 4.2

## 4. DISCUSSION

### 4.1 Macroscopic evolution of the physical parameters

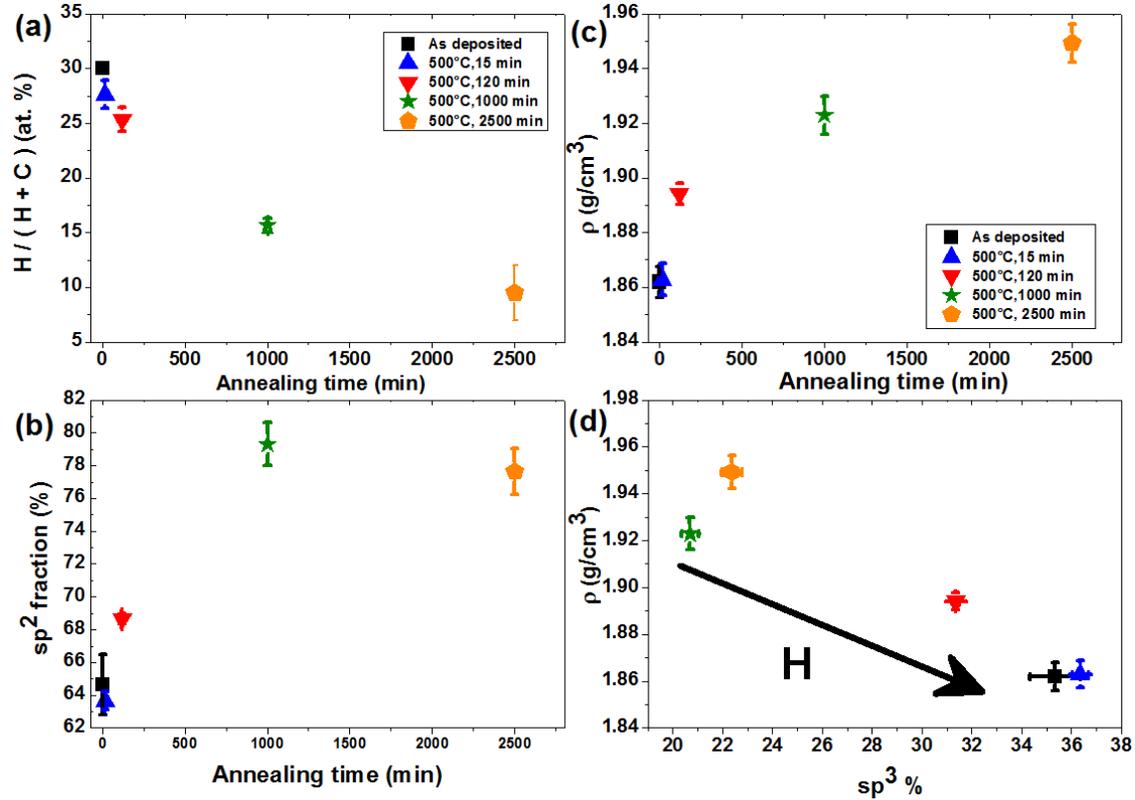

**Fig. 6** – Variation of (a) hydrogen content, measured by Raman spectroscopy (the error bars represent the dispersion of measures using the three laser wavelength) and (b) $sp^2$ fraction and determined by EELS with the annealing time. Variation of mass density (c) with annealing time (d) and with $sp^3$ fraction. (A color version of this figure can be viewed online)

The hydrogen content (H/(H+C)), determined from Raman spectroscopy, is shown in Fig. 6a. It is equal to 30 % for the as-deposited sample and then decreases to reach a value of nearly 9 % for the sample annealed during 2500 min. As already reported in the literature, the annealing of hydrogenated amorphous samples leads to hydrogen desorption from the thin films [87]. Fig. 6b shows the variation of the $sp^2$ fraction with the annealing time. This curve is divided in three steps: (*I*): between 0 and 15 min, (*II*): between 15 and 1000 min, and (*III*): between 1000 and 2500 minutes. For each of these sections, different processes, modifying the $sp^2$ fraction are taking place. Between 0 and 15 minutes, the value of the $sp^2$ fraction decreases from 64.7 ± 1.8

% to 63.7 ± 0.6 %. However, this variation is within the experimental error and is not statically significant due to the strong inhomogeneity of the as-deposited sample (section 4.2). Between 15 and 1000 minutes, the $sp^2$ fraction slowly increases from 63.7 ± 0.6 % to 79.3 ± 1.3 %. This is consistent with the weak ELNES variations discussed before. Furthermore, this fact highlights the subtle influence of isothermal annealing at 500°C, which triggers a slight but controllable conversion of $sp^3$ domains to $sp^2$ clusters. After a treatment of 2500 minutes, the $sp^2$ fraction is equal to 77.7 ± 1.4 %. Taking into account the experimental error, this value is close to the one at 1000 min. However it might also reflects the overall oxidation of the thin film as it will be discussed in the next section.

Fig. 6c shows the variation of the mass density with the annealing time, obtained from the low-loss EELS analyses. The value of the mass density starts from 1.862 ± 0.006 g.cm$^{-3}$ for the as-deposited samples and increase until reaching a value of 1.923 ± 0.007 g.cm$^{-3}$ for the sample annealed during 1000 minutes. The densification of the thin films is then consistent with the overall increase of the $sp^2$ fraction and hydrogen desorption, which are both observed during the step *II*. These values are also in good agreement with the values in literature of thin films with similar hydrogen content [3,88]. A further increase of the annealing time to 2500 minutes leads thus only to a slight increase of the mass density (1.949 ± 0.007 g.cm$^{-3}$). In fact, the increase of the plasmon energy (from 25.20 ± 0.04 eV to 25.36 ± 0.05 eV going from 1000 to 2500 min) in Eq. 1 is compensated by the high oxygen content ($X_O$ ~ 13.2 %, see section 4.2). However, it should be noted that this value has been calculated assuming an effective electron band mass which has been obtained from oxygen free-amorphous carbon [57,84] and then we cannot exclude some differences in these amorphous carbon with samples such high oxygen content. Fig. 6d shows the variation of the density with the $sp^3$ fraction. This plot has been used in the seminal work of Ferrari to distinguish the different kind of amorphous carbons [57]. For almost all the samples, the mass density decreases with the increasing of the $sp^3$ fraction and the decreasing of the H fraction. This behavior is the expected for a-C:H materials, while ta-C and ta-C:H materials follow the opposite trend, for which the mass density increases with the decreasing of the $sp^3$ fraction [57]. Two of the samples do not follow this behavior. The first one is the as-deposited sample, which contains the highest hydrogen fraction. This sample presents the lowest density but not the highest $sp^3$ fraction. The second sample is the sample annealed during 2500 min, which contains the lowest hydrogen fraction. This sample displays the highest density but not the lowest $sp^3$ fraction.

The deviations of these two samples from the expected behavior stresses the presence of other processes, which affect the $sp^2$ fraction and the mass density and which cannot only be explained by macroscopic observations.

In the following sections, closer attention will be devoted to ELNES inhomogeneites, which are related to variations of the local chemical environment. In particular, these ELNES analyses will be studied as a function of the distance between the electron probe and the substrate.

**4.2 In-depth chemical inhomogeneities**

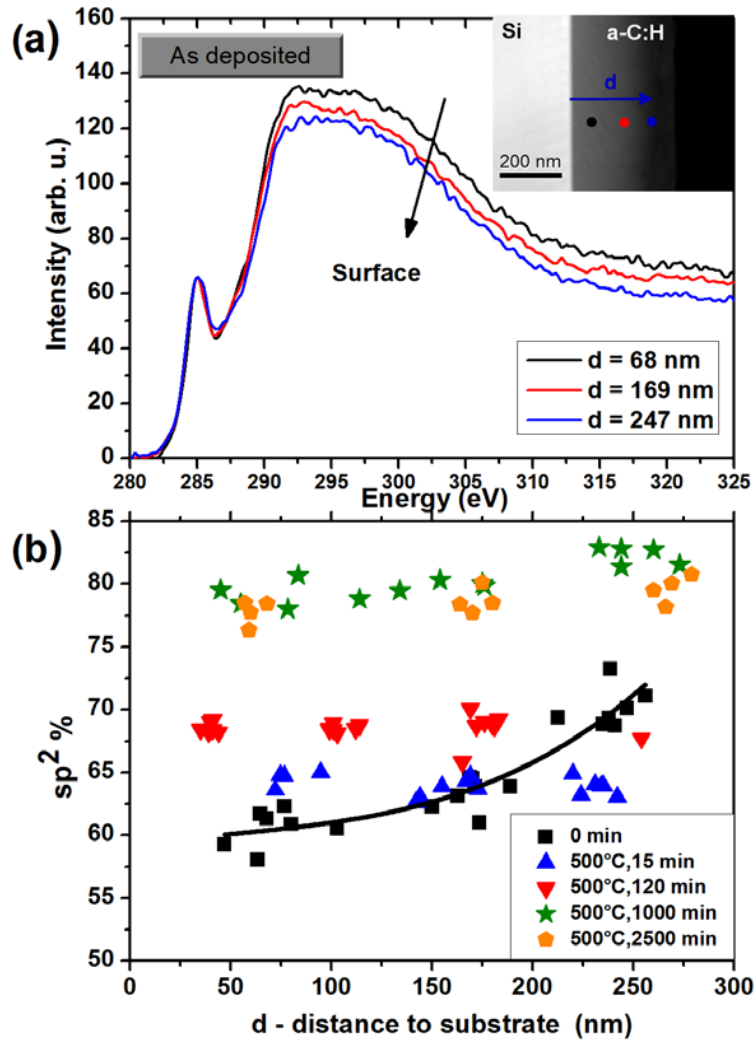

**Fig. 7** (a) Variation of the C-K ELNES of the as-deposited sample with the distance to the substrate. The inset shows a low-magnification STEM-HAADF micrograph of the as-deposited sample highlighting how the *d* parameter (distance between the electron probe and the substrate) is defined. (b) Variation of the sp$^2$ fraction, determined by EELS, as a function of the distance between the electron probe and the substrate. The black line is a guide for the eyes. (A color version of this figure can be viewed online)

Fig. 7a shows the ELNES of the as-deposited sample as a function of the distance between the substrate and the center of the electron probe (see the low-magnification image in the inset). A clear trend is observed from this figure: the intensity of the massif situated after 292 eV is lower for the spectra closer to the surface than for the spectra closer to the silicon substrate. For this sample, the $sp^2$ fraction is thus higher closer to the surface than for the spectra closer to the substrate. This result is quantitatively confirmed by the findings presented in Fig. 7b, which correspond to the variation of $sp^2$ fraction as a function of the distance to the silicon substrate for all the examined samples. For the as-deposited sample, the $sp^2$ fraction close to the substrate is equal to 59% and reaches 70 % close to the surface. The as-deposited sample presents then a strong gradient of $sp^2$ fraction and this is a unique characteristic comparing the other samples and thus cannot be ascribed to irradiation damages during sample preparation and STEM-EELS observation. Macroscopic dependence of the $sp^2$ fraction with the thickness of ultra-thin a-C:H and diamond-like thin films were already reported [89–91] and it was shown in particular that a-C:H thin films with a thickness inferior to 50 Å contain more $sp^2$ carbons [90]. Here, the opposite trend is observed, and then, another phenomenon is clearly taking place. This gradient could be induced by the deposition method and might be due to a gradient of the hydrogen fraction along the thin film or it could be induced by mechanical stress. This effect also explains why this sample could not be described by macroscopic observations. After 15 min, the annealing leads to the homogenization of the $sp^2$ fraction along the whole thin films and the $sp^2$ fraction increases then with the annealing time. This one illustrates directly the effect on the annealing on the local chemistry of the thin films. After 1000 minutes, the $sp^2$ fraction reaches the maximum value of 79 % and further annealing does not lead anymore to the conversion of $sp^3$ to $sp^2$ carbons. A slight difference close to the surface can be observed between 1000 and 2500 minutes, the $sp^2$ fraction being lower for the sample annealed 2500 minutes than for the one annealed 1000 minutes. This sole difference explains the macroscopic decrease of the $sp^2$ fraction reported in Fig. 6b for the sample annealed 2500 minutes. The sample annealed 25000 presents thus the same values of sp$^2$ fraction along the whole film. However and as it will be explained in the following, this behavior hides a decrease of the π*(C=C) contribution when going close to the surface.

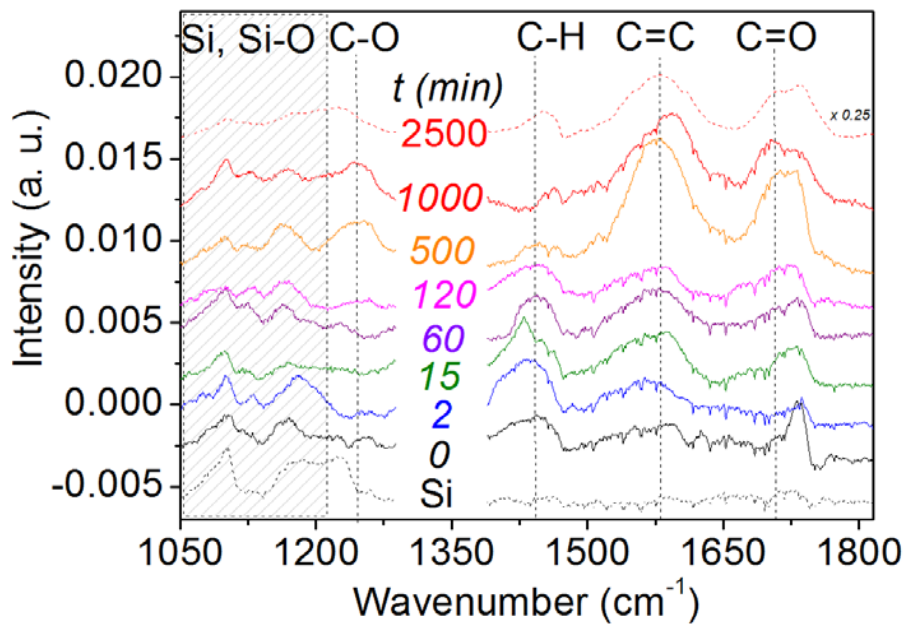

**Fig. 8** ATR spectra of the a-C:H samples heated at 500°C during 0 to 2500 minutes. The intensity of the sample annealed during 2500 minutes is divided by a factor four in the figure. (A color version of this figure can be viewed online).

Fig. 8 displays the infrared spectra recorded using the ATR setup. The silicon wafer on which were deposited the carbon layers give rise to signatures in the spectral window 1050-1240 cm$^{-1}$, thus limiting our interpretation of the carbon signatures falling in that range. On the other hand, no spectral features related to Si can be observed between 1400 and 1800 cm$^{-1}$. On the as-deposited sample, three bands are lying at 1442, 1580 and 1734 cm$^{-1}$. The first and second one are interpreted as due to C-H and C=C bonds, respectively while the third one is an artifact due baseline correction. When heating time increases, one can observe two new bands at 1248 and 1706 cm$^{-1}$ which are due to C-O and C=O bonds, respectively. Compared to the band related to C=C bonds, the intensity of the band related to C-H bonds decreases with time, which indicates a loss of hydrogen in the a-C:H, whereas the intensity of the bands related to C-O and C=O bonds increases with time, which indicates an oxidation of the a-C:H. This is particularly striking for the sample annealed during 2500 minutes.

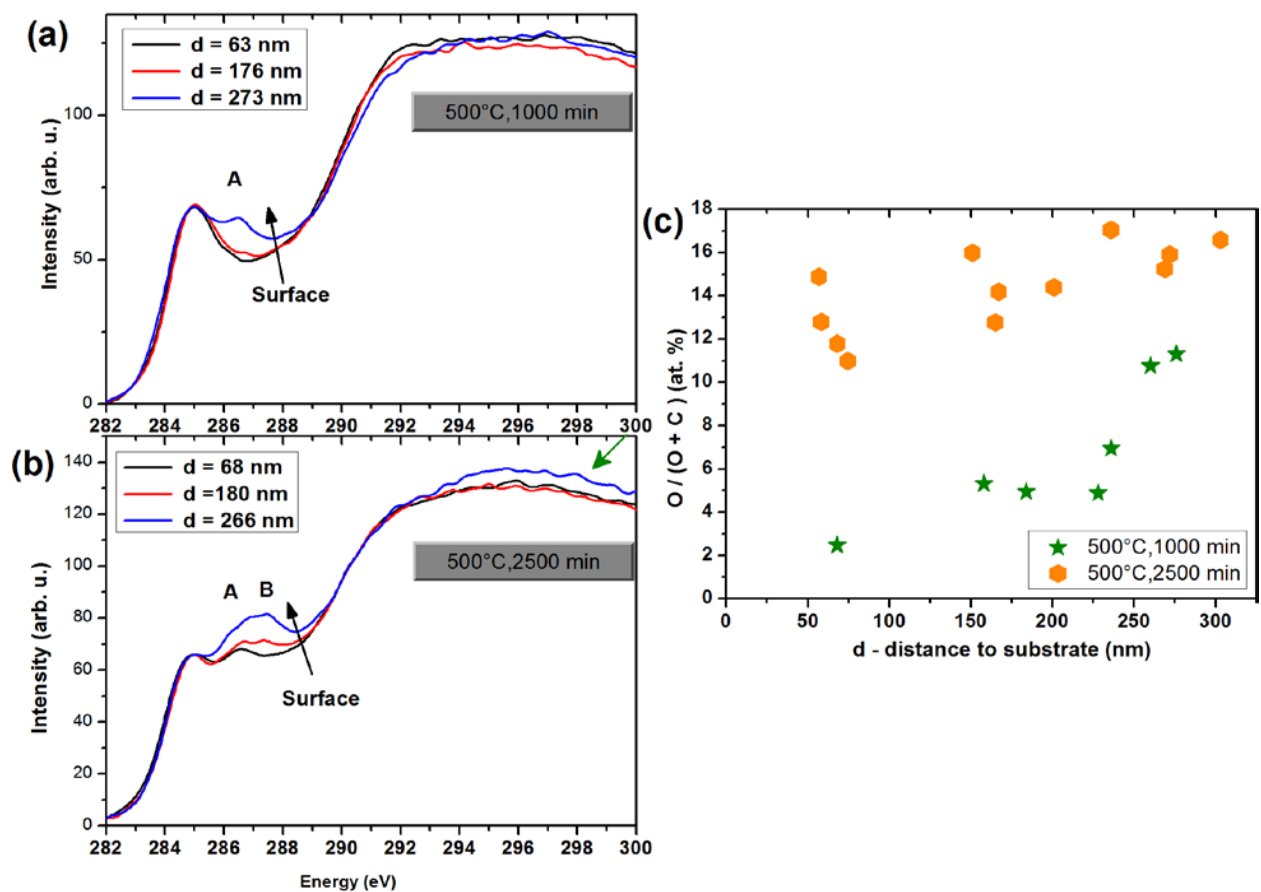

**Fig. 9** ELNES of the samples annealed during 1000 minutes (a) and 2500 minutes (b). For each spectrum, the distance between the electron probe and the substrate is indicated. The green arrow highlights the increase of the σ* contribution. (c) Oxygen content as a function of the distance between the electron probe and the substrate. (A color version of this figure can be viewed

Fig. 9a and 9b show the C-K edge ELNES of the samples annealed at 500°C during 1000 and 2500 minutes for several distances between the electron probe and the substrate, respectively. The sample annealed during 1000 minutes presents a supplementary peak (labeled *A* in Fig. 9a and is situated ~ 1.6 eV after the π* peak) which is only detected close to the surface (more than 250 nm from the substrate). However, this peak is always present in the spectra of the sample annealed during 2500 min, independently of the distance between the substrate and the electron probe. In addition, a supplementary peak (labeled *B* in Fig. 9b and situated ~ 2.4 eV after the π* peak) is highlighted only for the samples annealed 2500 minutes and only close to the surface. According to the literature, such supplementary peaks situated right after the π* peak might be ascribed to a plethora of transitions involving heterospecies (H and O in different bonding configurations…) and non-planar $sp^2$-bonded carbon (*i.e.* fullerene-like) [78,92–97]. For instance, aromatic C-OH group will give rises to transitions to π* states in the energy range 285.8-286.4 eV, C=O group to transitions to π* states in the energy range 286.5-288.3 eV, C-H group to transitions to σ* states in the energy range 287.6-288.2 eV and aromatic carbonyl and carboxyl groups to transitions to π* states in the energy range 287-288.5 eV [54,78,92–94,96]. The determination of the origins of these supplementary transitions based solely on the energy position is then particularly difficult. However, in our case, these transitions are only observed on the samples having the lowest hydrogen content and it is thus unlikely that these transitions could originate from C-H groups. They are more likely to be related to transitions to π* states. For the samples annealed 1000 and 2500 minutes, a slightly modified method has thus been used for the determination of the *R ratio* to take these contributions into account (Fig. S2 in SI). To get more insight on the oxygen content on these samples, we have performed core-loss EELS analyses with a larger energy dispersion (0.5 eV/channel) for recording simultaneously the C-K and O-K edges. The results of the O/(O+C) quantification as a function of the distance between the substrate and the electron probe is shown in Fig. 9c for the samples annealed during 1000 and 2500 min. As it can clearly be observed, the sample annealed during 1000 minutes is only oxidized at the surface (10% of O/(O+C) atomic fraction after 250 nm from the substrate), while the sample annealed during 2500 min is oxidized along the whole thin film. This observation matches well the behavior of the *A* peak. Thus, we can assign the peak *A* to C=O groups, and the peak *B* to aromatic carbonyl and/or carboxyl groups. These interpretations confirm previous observations performed on soot, oxidized nanodiamonds and other organic moieties [54,78,93]. The respective contributions of the C=C groups and carbo-oxygenated groups (inducing the *A* and *B* peaks) in the π* character is shown in the Tab. S1 in the Supplementary Data. These assignments allow us to estimate that ~ 6 % of the 79 % of $sp^2$ fraction of the sample annealed 1000 minutes corresponds to carbons linked to oxygen (the methodology to extract this information is detailed in the Supplementary Data). This effect is slightly more pronounced close to the surface and is followed by a slight increase of the C=C contribution. Thus, this explains the small increase of the total $sp^2$ fraction shown for this sample at the surface in Fig. 7b.

The sample annealed 2500 minutes, which shows a total $sp^2$ fraction close to 78 %, presents an important contribution of carbons linked to oxygen in the π* character (nearly 10 %). This sample shows a constant total $sp^2$ fraction along the thin film, however this is because the decrease of the C=C contribution at the surface is compensated by an increase of the contribution of carbons linked to oxygen. The increasing contribution of the peak *B* is thus followed by an increase of the σ* character as it can be shown on the EELS spectra (green arrow in Fig. 9b).

Finally, the non-negligible oxygen content of these two samples helps us to understand why they do not follow the expected variation of the mass density with the $sp^3$ fraction (Fig. 6d).

## 4.3 Validity of empirical-based Raman models

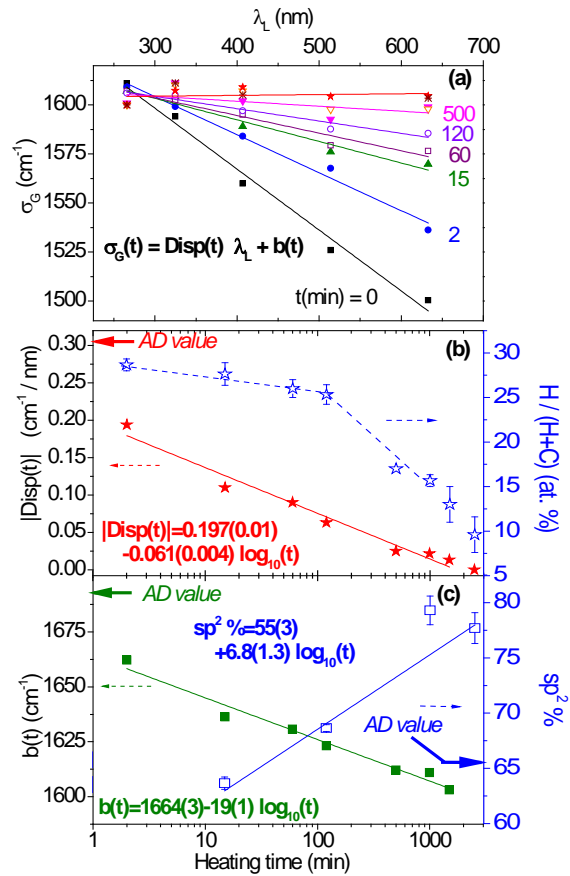

**Figure 10-** G band dispersion compared to EELS measurements. (a) G band dispersion for as deposited sample compared to heated samples (2 to 1500 minutes). Each curve has been linearly fitted. The slope (b) and the intercept (c) of the fit are displayed for all the samples. The H-content and $sp^2$ fraction (obtained by EELS) are also displayed, for comparison.

The EELS local analyses provide very rich information, allowing us to better understand and interpret the behavior of these materials. However, as abovementioned, multiwavelength Raman spectroscopy is a very powerful technique to determine structural and chemical properties of a given material, as it is a very fast method compared to other analytical techniques. As an example, an empirical relation linking the $sp^3$ content to the G band dispersion was derived for as-deposited amorphous carbons [29] which however fails to extract the $sp^3$ content for heated amorphous carbons. In addition, some care has to be taken in using formulae relating Raman spectroscopic parameters to structural and chemical properties: the spectroscopic parameters obtained by fitting depend strongly on the model used. In particular, there is an important limitation related to the fact that different kind of clusters building an amorphous carbons give spectral signatures overlapped in the same spectral range [30]. Recently Zhang *et al.* proposed a quantitative model showing that the dispersion of the G band, weighted by three types of clusters (nc-graphite, fused aromatic rings with size less than 2 nm and olefinic chains), is linearly related to the $sp^3$ content [30]. This model also supports our fittings and analytical method, consisting in taking the apparent value of the G band wavenumber, $\sigma_G$, without fitting. For the as deposited and heated samples (2 to 1500 minutes), the dispersion curves are displayed in Fig. 10a (the 2500 minutes is not shown but the dispersion is equal to zero). The G band wavenumber for all our samples vary as $\sigma_G = Disp\ \lambda_L + b$ (*Disp* being the slope and *b* the intercept) and the absolute value of the slope diminishes with the annealing time. From this plot and following the model of Zhang *et al.* [30], it is possible to determine the population of $sp^2$ carbon atoms embedded in aromatic clusters lower than 2 nm, in aromatic clusters higher than 2 nm and the in olefinic chains. This yields that the $sp^2$ carbon atoms are half embedded in aromatic clusters lower than 2 nm and half embedded in olefinic chains for the as-deposited sample, without carbons embedded in clusters with size higher than 2 nm. After 2 minutes at 500°C sample, only 10 % of $sp^2$ carbon atoms remain in aromatic clusters of less than 2 nm in diameter whereas roughly 40 % are now found in aromatic clusters with a size higher than 2 nm while the population of sp$^2$ carbon atoms embedded in olefinic chains remains unchanged. The fact that aromatic ring clusters are depleted earlier/faster than the olefinic chains is also supported by the presence of a band lying as a shoulder of the D band at 1200 cm$^{-1}$ (attributed to carbon chains, see Fig. 3), which remains in the spectrum even after 1000 minutes of heating at 500°C. However after 10 minutes of heating the fraction of carbons embedded in aromatic clusters lower than 2 nm reaches unrealistic negative values (-20 % for 120 minutes).

The evolution of *|Disp|* and *b* with the logarithm of the annealing time are displayed in Fig. 10b and 10c, respectively. They both follow a simple linear law with the logarithm of the annealing time, as suspected in [17]. Note that the R coefficient is 0.97 for both the linear fit of *|Disp|* and b. From Fig. 10c, similar trend is observed for the evolution of the sp$^2$ fraction with the logarithm of the annealing time. The following linear relation between *sp$^2$ %* and *|Disp|* can thus be determined:

$$sp^2\ \% = 74.6 - 86.9\ |Disp|/(cm^{-1}/nm)$$

A similar linear relationship has already been reported, but with different coefficients (*sp$^2$ % = 100 - 230 |Disp|/(cm$^{-1}$/nm)* in [30]). Applying this relation to our samples will lead to overestimate the sp$^2$ % when compared to the values obtained by EELS (75% instead of 63% for 15 minutes, 85% instead of 68% for 120 minutes for example). In Fig. 10b, it is noteworthy that the H content does not follow a linear law with $\log_{10}(t)$ in the total range of heating times but displays a change of slope between 100 and 500 minutes, the hydrogen release being faster for t > 500 minutes. For heating time lower than 120 minutes the H content evolves with log10(t) following :

$$H/(H+C)\ (at.\ \%) = 29.4\ (0.4) - 1.90\ (0.26)\ \log_{10}(t)$$

A similar relationship can be derived from [30] with the same slope but with an intercept 20% lower ($H/(H+C)$ (at. %) = 22.8 (0.5) -1.92 (0.31) $log_{10}(t)$). It thus can stated that our work and the model of Zhang et al. [30] agree qualitatively (same trends for the different laws), but there are still some quantitative disagreement. In particular, the population of $sp^2$ carbon atoms does not make any sense after 120 min (negative values) nor does the $sp^2$ fraction when compared to the EELS results or the determination of the hydrogen content. Even by excluding the samples showing the higher oxygen content (1000 and 2500 minutes), the discrepancy remains. This could highlight than even a slight oxidation might have a strong influence on the Raman parameters (an increase of the bands related to infrared C=O bonds is observed between 120 and 500 minutes). Other possibilities might involve the different methods used to determine the hydrogen content and the $sp^2$ fraction or the fact fail to reproduce the thermal behavior of our samples which are better organized.

## 4. CONCLUSION

Hydrogenated amorphous carbon thin films on silicon substrate were submitted to an isothermal annealing up to 2500 minutes and investigated by a combination of Raman, ATR and EELS spectroscopies. In addition, a method to determine the $sp^2$ fraction from the EELS spectra has been proposed, which does not require specific experimental setup and can include contributions from heterospecies. This fitting and analytical method shed light on the processes taking place during the thermal treatments as well as on the physical and chemical properties of the samples. A complete understanding of the local chemistry of the thin films has been made and it shed the light on the high complexity of such samples. Surprisingly, strong in-depth inhomogenities of the local chemistry has been highlighted for several samples. The as-deposited sample presents a strong gradient of the $sp^2$ fraction, which is higher at the surface of thin film. An annealing of 15 minutes is enough to lead to the homogenization of the thin films and further annealing time leads a slow conversion of the $sp^3$ domains to $sp^2$ clusters followed by hydrogen desorption of the thin films. Starting from 1000 minutes of annealing, a slight oxidation of the whole thin film is measured at the surface of the thin film which induced the presence of another $\pi^*$ contribution linked to C=O groups. Further annealing at 2500 minutes leads to an increase of the oxidation of the whole thin film and to the appearance of another $\pi^*$ contribution linked to aromatic carbonyl and/or carboxyl groups which is also followed by an increase of the $\sigma^*$ character. In addition, the possibility to extract directly quantitative information on the local chemistry solely based on the variation of Raman parameters has been investigated. Even by excluding the sample showing the higher oxidation, our work does not validate previous model of the literature used to determine the $sp^2$ fraction and the hydrogen content. This clearly illustrates the importance of combining several techniques to extract properly and reliably this kind of information.

**ACKOWLEDGMENTS**

The TEM and EELS studies were conducted at the Laboratorio de Microscopias Avanzadas, Instituto de Nanociencia de Aragon, Universidad de Zaragoza, Spain. Some of the research leading to these results has received funding from the European Union Seventh Framework Programme under Grant Agreements 312483- ESTEEM2 (Integrated Infrastructure Initiative – I3) and 604391 Graphene Flagship, as well as from EU H2020 Grant Agreement 696656 Graphene Flagship. R.A. acknowledges funding from the Spanish Ministerio de Economia y Competitividad (FIS2013-46159-C3-3-P) and from EU H2020 ETN project "Enabling

Excellence" Grant Agreement 642742. The support of the Australian Research Council's Linkage Infrastructure, Equipment and Facilities Program (LE130100119) is also gratefully acknowledged for funding the Horiba Labram Raman system.

**Supplementary Data**

Modified method used for the determination of the *R* ratio taking into account the contribution of hererospecies. Low-loss EELS spectrum showing the Drude model fit used to extract the plasmon energy. Calibration curve between $H_D/H_G$ Raman parameters used in this paper to derive the H content. Respective contributions of the C=C groups and carbo-oxygenated groups (related to A and B peaks) in the $\pi^*$ character for the samples annealed 1000 and 2500 minutes.

# Supplementary Data for

# Advanced Spectroscopic Analyses on a:C-H Materials: Revisiting the EELS Characterization and its Coupling with multi-wavelength Raman Spectroscopy

L. Lajaunie[1‡], C. Pardanaud[2‡], C. Martin[2], P. Puech[3], C. Hu[4,5], M. J. Biggs[4,5], R. Arenal[1,6*]

[1]*Laboratorio de Microscopias Avanzadas (LMA), Instituto de Nanociencia de Aragon, Universidad de Zaragoza, 50018 Zaragoza, Spain*

[2]*Aix-Marseille Université, CNRS, PIIM, Marseille, France*

[3]*CEMES, Toulouse, France*

[4]*School of Chemical Engineering, The University of Adelaide, SA 5005, Australia*

[5]*School of Science, Loughborough University, Loughborough, LE11 3TU, UK*

[6]*Fundacion ARAID, 50018 Zaragoza, Spain*

\* *Corresponding author:* Tel: +34 976 762 985, E-mail address: arenal@unizar.es

[‡] These authors contributed equally

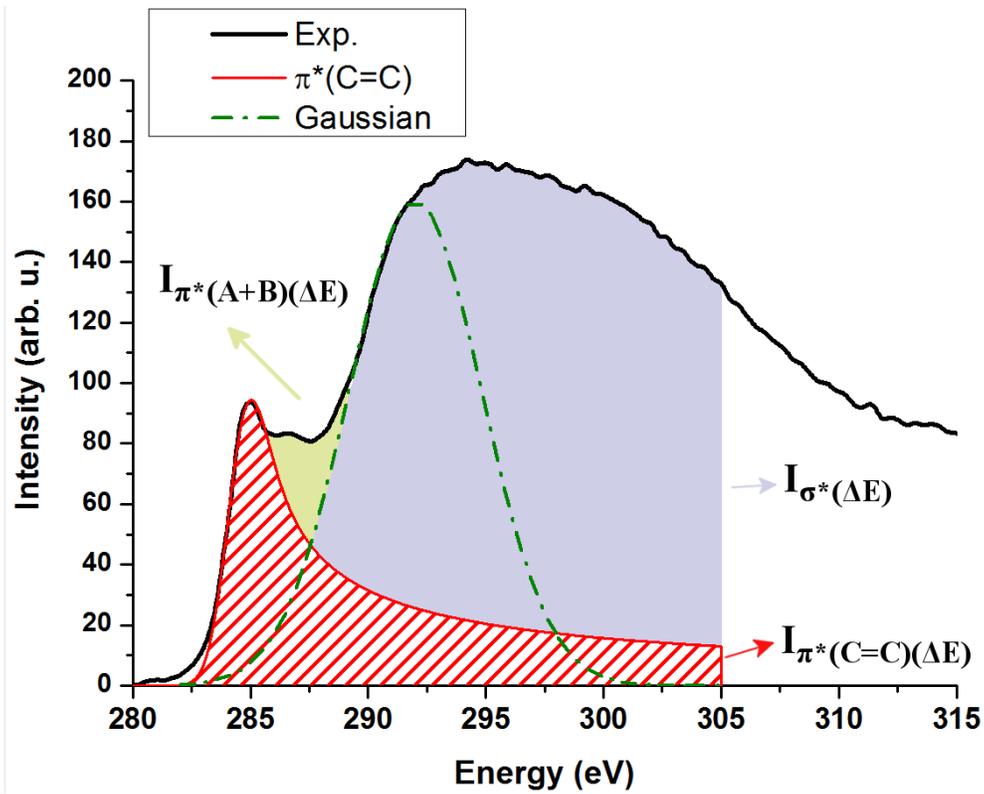

**Fig. S1** Modified method used for the determination of the *R* ratio for the samples annealed at 1000 and 25000 minutes.

Fig. S1 shows the modified method used in for the determination of the *R* ratio for the samples annealed at 1000 and 2500 minutes. The spectral weight of the contributions π*(C=C) is still determined by using a sum of Gaussian functions of equal standard deviation (red hatched area in Fig. S1). However, a Gaussian function centered at 292 eV (green dot curve in Fig. S1) is introduced to reproduce the beginning of the σ* massif (blue area in Fig. S1). The spectral weight of the peaks A and B, π*(A+B) (yellow area in Fig. S1), which are related to the oxygenation of the thin films is then determined by the intersection between this Gaussian function centered at 292 eV and the π*(C=C) contribution. Finally the total spectral weight, $I_{\pi^*+\sigma^*(\Delta E)}$, is determined by area integration of the experimental curve up to 305 eV. The R ratio is then given by $R = (I_{\pi^*(C=C)(\Delta E)} + I_{\pi^*(A+B)(\Delta E)}) / I_{\pi^*+\sigma^*(\Delta E)}$.

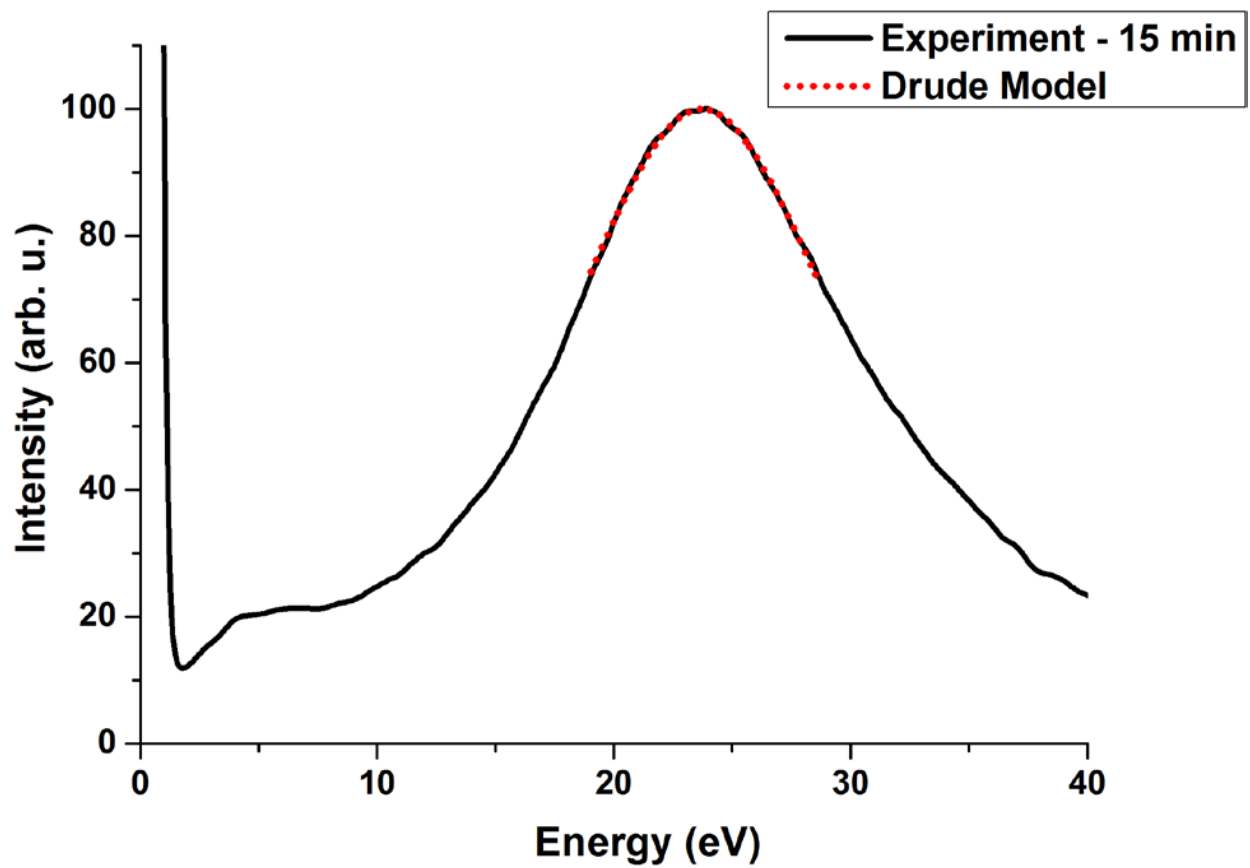

**Fig. S2** Low-loss EELS spectrum of the sample annealed 15 min. The red dotted curve corresponds to the Drude model fit used to extract the plasmon energy, $E_P$.

Fig. S2 shows the low-loss EELS spectrum of the samples annealed 15 min. The experimental curve (black line) has been fitted by the Drude model (red dotted line) to extract the plasmon energy, $E_P$, which has been used to estimate the mass density.

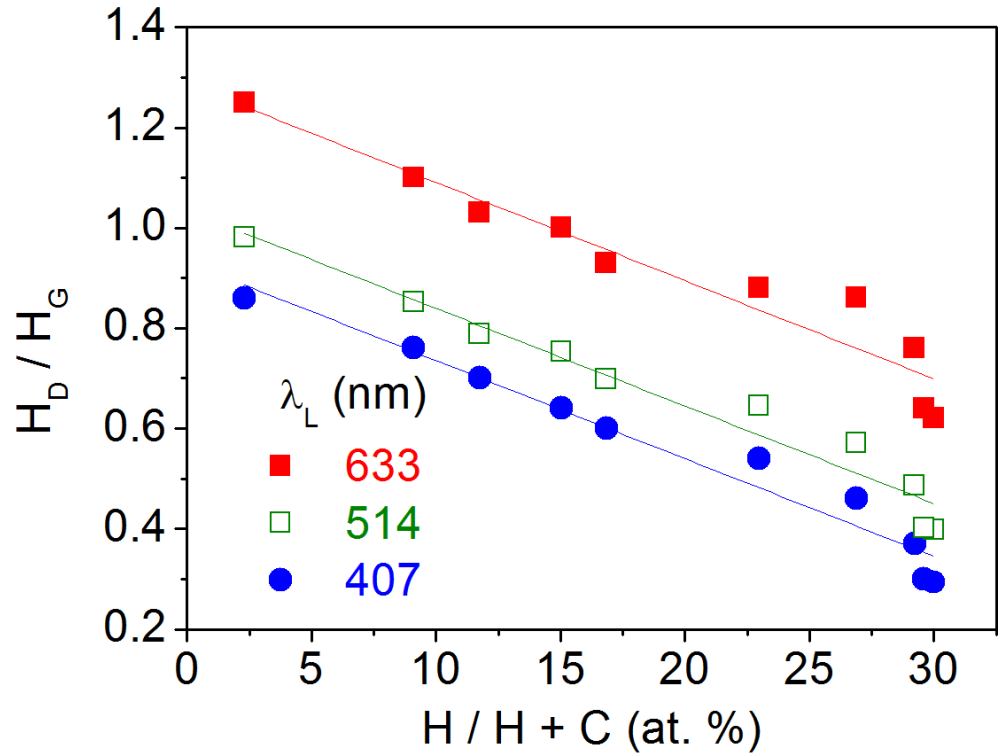

**Fig. S3** Calibration curve between $H_D/H_G$ (Raman spectroscopic parameter) and the H content measured by ion beam analysis under ultrahigh vacuum.

The first aim of using Raman spectroscopy in this paper was to determine the H-content, following the method presented in [1], using the fact that $H_D/H_G$ evolves linearly with the H-content for post-heated samples. The method, originally published using $\lambda_L$=514 nm is extended here to 633 and 407 nm. Fig. S3 displays $H_D/H_G$ evolution in function of hydrogen content of samples heated under ultra-high vacuum and then measured by ion beam analysis (see [1] and references therein for details). For 633, 514 and 407 nm $H_D/H_G$ increases linearly as the H content decreases, with the same slope. Only the intercept changes with the wavelength. The behavior changes for 325 and 266 nm (not shown here). This could be due to a change in the G band Raman cross section of a-C:H using UV laser [2]. For the samples heated in argon atmosphere at 500°C from 2 to 2500 minutes, we took the corresponding $H_D/H_G$ values presented in Fig. 4c and applied the linear relations displayed in Fig. S1 to deduce the H-content.

|  |  | Whole sample | d ~ 60 nm | d ~ 160 nm | d ~ 260 nm |
|---|---|---|---|---|---|
| **1000 minutes** | π*(C=C) | 73 % | 72 % | 73 % | 75 % |
|  | π*(A+B) | 6 % | 5 % | 6 % | 8 % |
|  | Total sp$^2$ | 79 % | 77 % | 79 % | 83 % |
| **2500 minutes** | π*(C=C) | 67 % | 70 % | 68 % | 66 % |
|  | π*(A+B) | 11 % | 9 % | 11 % | 12 % |
|  | Total sp$^2$ | 78 % | 79 % | 79 % | 78 % |

**Tab. S1** Respective contributions of the C=C groups and carbo-oxygenated groups (related to A and B peaks) in the π* character for the samples annealed 1000 and 2500 minutes. The values are given for the whole samples as well as a function of the distance between the electron probe and the substrate (*d* parameter). The total sp$^2$ fraction is given by π*(C=C) + π*(A+B).